\documentclass[preprint,review,3p,times]{elsarticle}

\usepackage[utf8]{inputenc}

\usepackage{amsmath}
\usepackage{amssymb}

\usepackage{graphicx}
\usepackage{tabularx}
\usepackage{threeparttable}

\usepackage{caption}
\usepackage{subcaption}
\usepackage[hidelinks]{hyperref}
\usepackage{xcolor}
\usepackage[normalem]{ulem}

\title{MEMS-based \textit{in situ} electron-microscopy investigation of rapid solidification and heat treatment on eutectic Al-Cu}

\author[add1]{Phillip Dumitraschkewitz}
\ead{phillip.dumitraschkewitz@unileoben.ac.at}
\author[add3]{Matheus A. Tunes}
\ead{matheus.tunes@unileoben.ac.at}
\author[add1]{Cameron R. Quick}
\author[add1]{Diego Santa Rosa Coradini}
\author[add1]{Thomas M. Kremmer}
\author[add2]{Parthiban Ramasamy}
\author[add1]{Peter J. Uggowitzer}
\author[add1]{Stefan Pogatscher}
\ead{stefan.pogatscher@unileoben.ac.at}

\address[add1]{Chair of Nonferrous Metallurgy, Department of Metallurgy, Montanuniversitaet Leoben, Franz-Josef-Str. 18, 8700, Leoben, Austria}
\address[add2]{Erich Schmid Institute of Materials Science, Austrian Academy of Sciences, Jahnstraße 12, A 8700 Leoben, Austria}
\address[add3]{Materials Science and Technology Division, Los Alamos National Laboratory, United States.}

\date{\today}

\begin{document}
\begin{abstract}
The solidification behavior of a eutectic AlCu specimen is investigated via \textit{in situ} scanning transmission electron microscope (STEM) experiments. Solidification conditions are varied by imposing various cooling conditions via a micro-electro-mechanical system (MEMS) based membrane. The methodology allows the use of material processed by a melting and casting route close to industrial metallurgically fabricated material for \textit{in situ} STEM solidification studies. Different rapid solidification morphologies could be obtained solely on a single specimen by the demonstrated strategy. Additional post-solidification heat treatments are investigated in terms of observation of spheroidization of lamellas during annealing at elevated temperatures.
\end{abstract}
\maketitle

\section{Introduction}
\noindent Eutectic alloys are classical and well studied metallurgical systems. In particular, the Al-Cu system is a regular eutectic system, which is defined by coupled growth of phase constituents from the melt ~\cite{Glicksman.2011, Kurz.}.

The rise of additive manufacturing~\cite{Gorsse.2017} and rapid solidification~\cite{W.Kurz.1994, Glicksman.2011, Kurz.2019}, as well as the emergence of the so-called eutectic high-entropy alloys (EHEA)~\cite{Chanda.2020, Shi.2019}, has led to new fundamental research efforts in the scope of eutectic alloys which resulted in a revisit to the Al-Cu system~\cite{Lei.2017, McKeown.2014, McKeown.2016, Bathula.2020}.

Regular eutectic systems are characterized by a lamellar or rod morphology, which is the result of the interfacial $\alpha$ factor for the constituent phases. The interfacial factor mainly depends on the entropy of fusion and on crystal structure and orientation. If both eutectic constituents have an interfacial factors $\alpha \leq 2$, a regular eutectic is expected. Depending on the composition of the near eutectic binary, either rod or lamellar morphology is expected.~\cite{Glicksman.2011}

Tiller~\cite{tiller1958liquid} formulated scaling laws for the dynamics of lamellar eutectic growth from the melt (eutectic scaling laws, see Equations~\ref{eq:tiller_1},~\ref{eq:tiller_2} and~\ref{eq:jackson_hunt}) {by using the principle of a minimum of the total interfacial undercooling $\Delta T$}. The eutectic scaling laws describe the relationship between undercooling, solidification velocity $v$ and resulting {optimal steady state} lamellar spacing $\lambda$. Tiller's model was later further refined by Hunt and Jackson~\cite{K.A.JacksonandJ.D.Hunt.}.~\cite{Glicksman.2011, Lemaignan.1981}

\begin{align}
    \label{eq:tiller_1}
    \lambda^2 v  &= \text{const.}\\
    \label{eq:tiller_2}
    \frac{\Delta T^2}{v}  &= \text{const.}\\
    \label{eq:jackson_hunt}
    \Delta  T\lambda  &= \text{const.}
\end{align}

Under high solidification velocities, as common for rapid solidification processing (RSP), the theoretical model defined by the eutectic scaling laws has limited applicability since different microstructures are able to form such as degenerate eutectics, cell/dendrites, bands and extended solid solutions~\cite{S.C.GILLW.KURZ.1993}. 

Gill and Kurz~\cite{S.C.GILLW.KURZ.1993, S.C.GILLW.KURZ.1995} experimentally and theoretically investigated a microstructure selection map for the Al-Cu system. \textit{Ex situ} experiments within a transmission electron microscope (TEM) on rapid laser solidification processed material were conducted~\cite{S.C.GILLW.KURZ.1993} and compared to predictive theoretical calculations using eutectic, dendritic, banding and plane front growth models~\cite{S.C.GILLW.KURZ.1995}. The morphology transition for the eutectic composition, from low to highest experimental solidification velocities, is reported in Equation~\ref{eq:sequence}~\cite{S.C.GILLW.KURZ.1995}.

\begin{align}
    \label{eq:sequence}
    \text{lamellar eutectic}\rightarrow \text{cellular and dendritic}\rightarrow\text{banded}
\end{align}

In general, ultimately increasing the solidification velocity can lead to partitionless solidification via solute trapping resulting in a partition coefficient of unity~\cite{MICHAELJ.AZIZTHEODOREKAPLAN.1988}, the formation of quasi-crystals~\cite{Kurtuldu.2018} or vitrification as experienced in bulk metallic glasses~\cite{Loffler.2003}.

For Al-Cu alloys, innovative methods have permitted direct and real-time observation of the solidification process in hypo-eutectic compositions in recent years~\cite{McKeown.2014,McKeown.2016,Bathula.2020}. By means of dynamic TEM (DTEM)~\cite{McKeown.2014} and movie-mode TEM (MM-TEM)~\cite{McKeown.2016,Bathula.2020}, the solidification behavior of an \textit{in situ} pulse-laser-melted pre-deposited hypo-eutectic Al-Cu film has been investigated. The material was prepared by electron beam evaporation of the pure materials onto a Si$_3$N$_4$ membrane. With a pulsed laser, an elliptical melt pool of $\approx 50$ $\mu$m was created locally and solidified by natural cooling, mainly driven by in-plane heat conduction of the surrounding solid. The solidification velocity increased during the solidification and reached a maximum of $\approx$ 1.4 m/s.~\cite{Bathula.2020}

A different approach for \textit{in situ} electron-microscopy rapid solidification studies in the Al-Cu eutectic system is herein investigated. In general, the methodology presented allows usage of material processed by a melting and casting route, which is closer to industrial, metallurgically fabricated material than usual in \textit{in situ} S/TEM solidification studies. { Moreover, the sample production method is expected not to be limited to a single alloy system. Additionally, the method can be time-saving if compared to a focused ion beam (FIB) sample production routine. The applicable time-temperature (t-T) programs for heat treatments are various below the maximum temperature range 1200 \textsuperscript{o}C~\cite{Allard.2009} of the membrane. This is especially true for relative short time spans, and could also be used to mimic t-T profiles of additive manufacturing cycles including solidification.}

{Utilizing a MEMS-based heating/cooling membrane, an electron transparent specimen, prepared by a simple method, is investigated. To demonstrate the capabilities of the methodology, different solidification velocities are explored in a single sample by imposing different cooling conditions via the chip.} Additional post-solidification heat treatments are conducted and changes on the microstructure in terms of spheroidization of lamellas are investigated.

\section{Experimental methods}
The material was produced by induction melting (Indutherm MC100V) and die casting, starting  from the pure metals Al (99.99 wt.\%) and Cu (99.99 wt.\%) to a target nominal composition of 17.39 at.\% Cu. The melting process was conducted under Ar atmosphere. The material was melted at 700 $^\text{o}$C, held for approximately 10 min, and cast.

Pieces of the ingot were cut, ground and analyzed via optical emission spectroscopy (OES) of type SPECTROMAXx. The composition measured was 82.48 at.\% Al and 17.52 at.\% Cu.

A volume of approx. $2\times 10^3$ mm, cut from the bottom third of the ingot, was used for melt-spinning. Several meters of ribbons could be produced of usable quality. The thickness of the produced ribbons varied from approximately 30 $\mu$m to 50 $\mu$m.

For scanning electron microscopy (SEM) analysis a SEM type Jeol JSM-IT300 equipped with a EDS system (Oxford X-Max\textsuperscript{50}) was used.

For STEM sample preparation, pieces of the melt-spun ribbons were cut and manually polished. The polished ribbons were punched and electro-polished with a mixture of a 1/3 nitric acid (HNO$_3$) and 2/3 methanol (CH$_3$OH). A jet electro-polishing (JEP) setup was used (TenuPol-5). The electrolyte was cooled down with LN$_2$ to -20 $^\text{o}$C and a voltage of 12 V was applied. After JEP, the material was washed in a sequence of three beakers containing pure methanol.

Following this, a small piece of material ($\approx 50$ $\mu$m) was cut from the electro-polished sample with a scalpel. The small sample was positioned on a Protochips Fusion Select \textit{in-situ} heating/cooling holder {chip} by hand using a natural grown animal hair as a manipulator stylus (for more details on this new procedure please see Reference~\cite{Tunes.2020,tunes2020contaminationfree}). The entire positioning and cutting was performed using a stereo microscope. This procedure is initially known from chip calorimetry~\cite{MettlerToledoGmbH.,Yang.2016,Shamim.2014,Poel.2012}, though the samples used there are comparatively thicker, usually in the range of several $\mu$m. The sample was positioned such that the thin, electro-transparent area covered the MEMS membrane holes which are intended for observation of a sample. Due to the size of the sample, several membrane holes were fully or partly covered, see {exemplary} Figure~\ref{fig:sketch_membrane}.

{For temperature control the Protochips Fusion Clarity program was used, which operates a Keithley 2450, utilizing a standard control time-step time of 100 ms.}

The sample was investigated by scanning transmission microscopy high angle annular dark field (STEM-HAADF) and EDS with a ThermoFisher Scientific\textsuperscript{TM} Talos F200X G2 scanning transmission electron microscope with an acceleration voltage of 200 kV {and at a pressure of $\approx 8 \times 10^{-6}$ Pa.}

\section{Results}
\begin{figure*}
    \begin{subfigure}[c]{0.91\linewidth}
        \begin{subfigure}[t]{\linewidth}
        \centering
            \begin{subfigure}[t]{0.32\linewidth}
                \includegraphics[width=\linewidth]{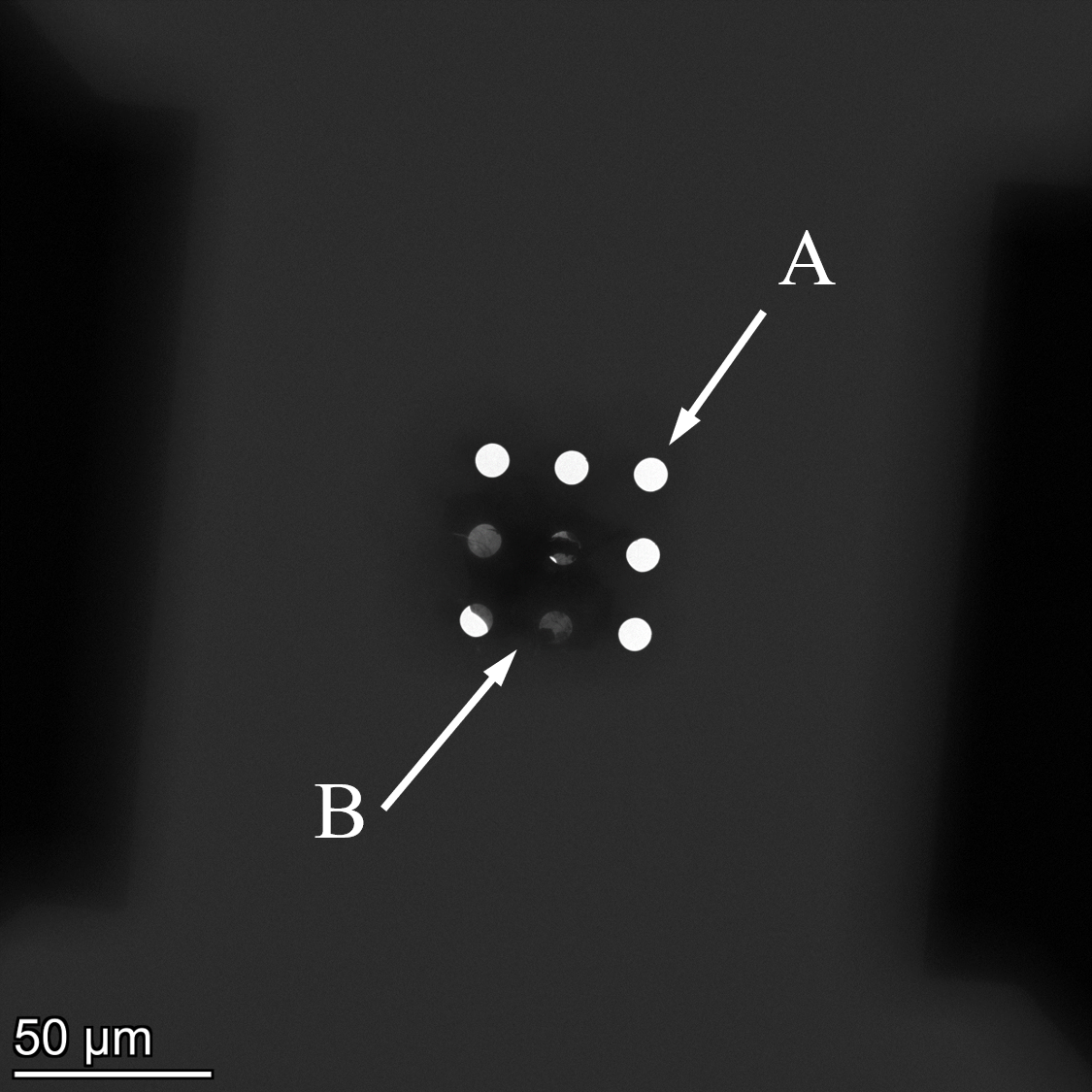}
                \caption{Membrane overview}
                \label{fig:sketch_membrane}
            \end{subfigure}
            \begin{subfigure}[t]{0.32\linewidth}
                \includegraphics[width=\linewidth]{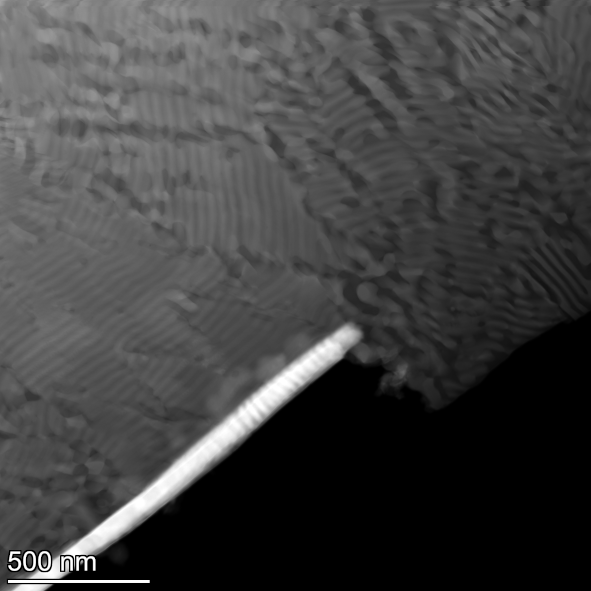}
                \caption{Pristine specimen}
                \label{fig:pristine_intro}
            \end{subfigure}
            \begin{subfigure}[t]{0.32\linewidth}
                \includegraphics[width=\linewidth]{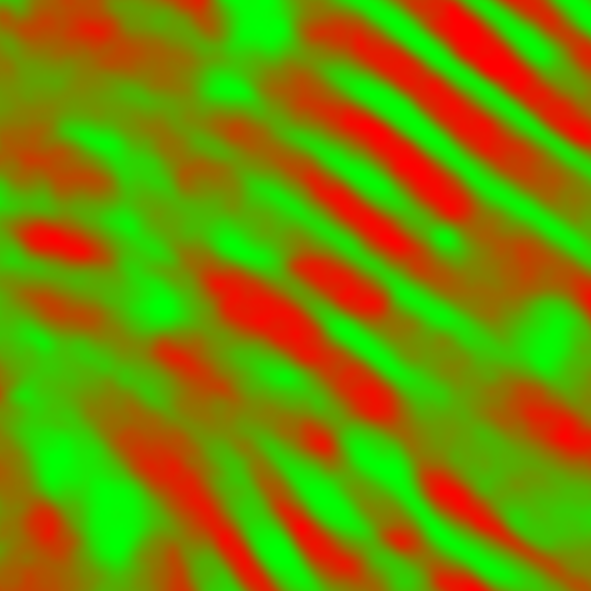}
                \caption{EDS pristine}
                \label{fig:EDS_pristine}
            \end{subfigure}
        \end{subfigure}
        \begin{subfigure}[c]{\linewidth}
        \centering
            \begin{subfigure}[t]{0.32\linewidth}
                \includegraphics[width=\linewidth]{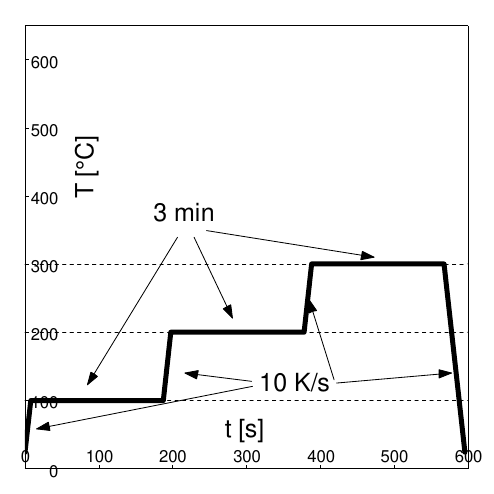}
                \caption{Coarsening temperature program}
                \label{fig:t_prog_1}
            \end{subfigure}
            \begin{subfigure}[t]{0.32\linewidth}
                \includegraphics[width=\linewidth]{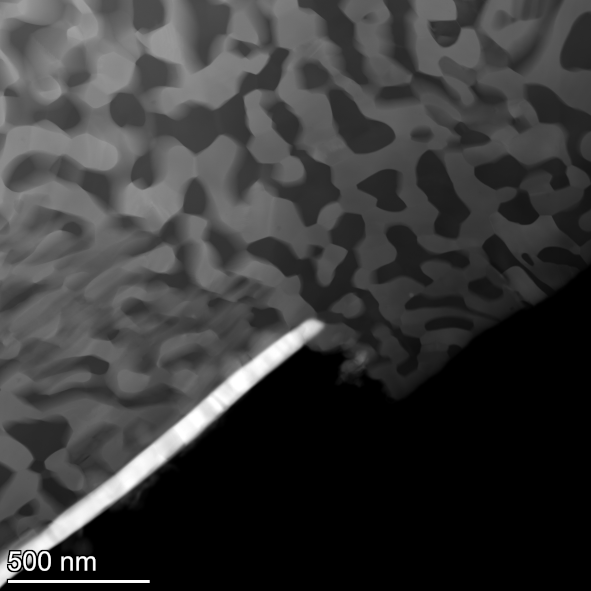}
                \caption{3 min at 300 $^o$C}
                \label{fig:exp_1_intro}
            \end{subfigure}
            \begin{subfigure}[t]{0.32\linewidth}
                \includegraphics[width=\linewidth]{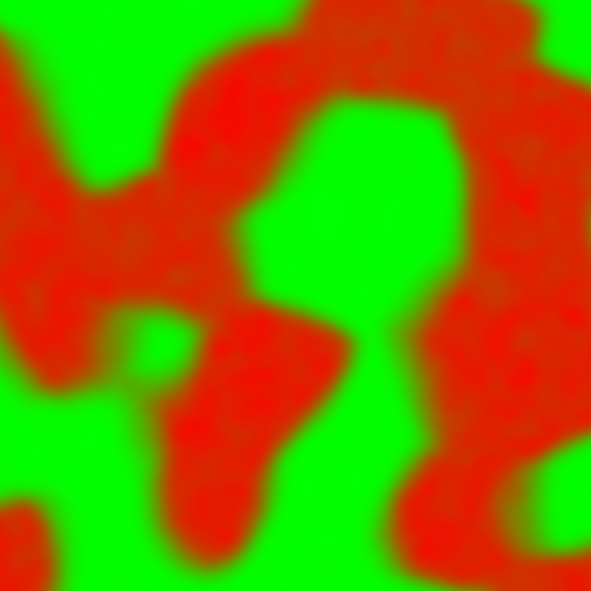}
                \caption{EDS coarsened}
                \label{fig:exp_1_EDS}
            \end{subfigure}
        \end{subfigure}
        \begin{subfigure}[c]{\linewidth}
        \centering
            \begin{subfigure}[t]{0.32\linewidth}
                \includegraphics[width=\linewidth]{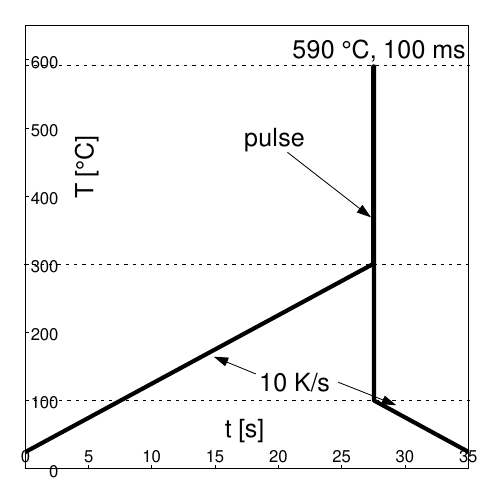}
                \caption{Temperature program re-solidification}
                \label{fig:t_prog_rest}
            \end{subfigure}
            \begin{subfigure}[t]{0.32\linewidth}
                \includegraphics[width=\linewidth]{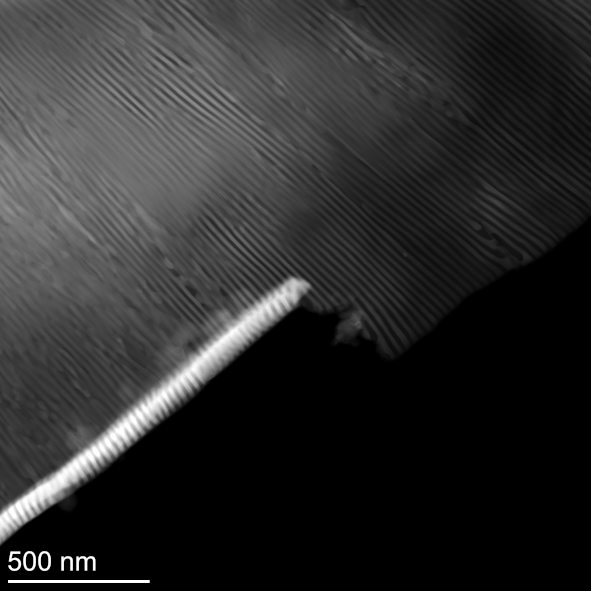}
                \caption{Re-solidified}
                \label{fig:exp_12_intro}
            \end{subfigure}
            \begin{subfigure}[t]{0.32\linewidth}
                \includegraphics[width=\linewidth]{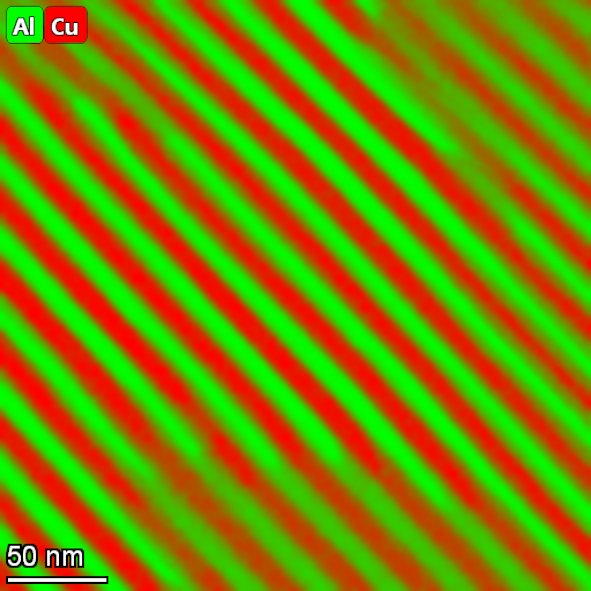}
                \caption{EDS re-solidified}
                \label{fig:exp_12_EDS}
            \end{subfigure}
        \end{subfigure}
    \end{subfigure}
    \begin{subfigure}[c]{0.05\linewidth}
        \includegraphics[width=\linewidth]{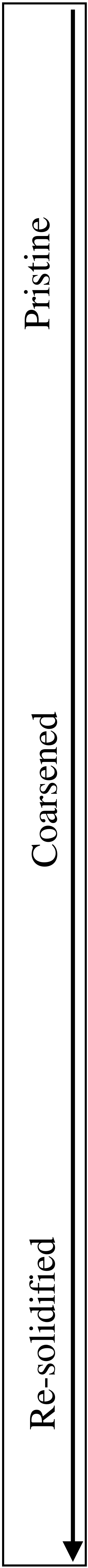}
    \end{subfigure}
\caption{\label{fig:pristine} Overview of the membrane, HAADF and EDS images of the pristine, coarsened pristine and re-solidifed material with corresponding temperature programs. a) Membrane overview with membrane holes~\cite{tunes2020contaminationfree} (feature A) and specimen (feature B). Following microstructures are obtained: b,c) lamellar colonies with varying orientation; e,f) interconnected spheroidized grains; h,i) unidirectional lamellas. Bright areas in HAADF show the $\theta$-Al$_2$Cu and dark areas are $\alpha$-Al phase as identified by EDS.  h,i) After re-solidification a newly formed nanostructured hierachy is obtained, consisting of $\alpha$-Al and $\theta$-Al$_2$Cu unidirectional lamellas. The bright feature, reaching approximately into the center of the figure b,e,h), is a roll-up of the sample.}
\end{figure*}

{In the following, the morphology for different conditions resulting from different time temperature programs are presented. In general, the results are organized according to the structure of the overview Figure~\ref{fig:pristine}.} Figure~\ref{fig:pristine_intro} and~\ref{fig:EDS_pristine} show the pristine sample. In Figure~\ref{fig:t_prog_1},~\ref{fig:exp_1_intro} and~\ref{fig:exp_1_EDS} the time temperature program and the coarsened state are reported; furthermore, in Figure~\ref{fig:t_prog_rest},~\ref{fig:exp_12_intro} and~\ref{fig:exp_12_EDS} the respective information and images of a melted and re-solified state are given.

The morphology of the pristine material and the specimen is briefly described in the following last paragraph due to the generality of observed features, which is also referred to in later sections.

{We give two detailed examples for application of our newly developed methodology. Firstly, the description of the spheroidization behavior, especially of a lamellar structure, including the measurement of an interface velocity follows in Section~\ref{sec:res_spheroidization}. And secondly, structures generated upon melting and re-solidification are presented in Section~\ref{sec:res_melting}.}

In Figure~\ref{fig:pristine_intro}, a high-angle annular darf field (HAADF) image of the pristine, as-meltspun, sample is shown. No unidirectional morphology of lamellas is obtained in the field-of-view, but colonies of lamellas can be identified in subdomains. The dark areas are identified as the $\alpha$-Al eutectic constituent, due to the z-contrast of HAADF and the relative high z-number of Cu, and the bright areas accordingly $\theta$-Al$_2$Cu. Energy dispersive X-ray spectroscopy (EDS, seen in Figures~\ref{fig:EDS_pristine},~\ref{fig:exp_1_EDS} and~\ref{fig:exp_12_EDS}) does confirm the high Cu content of the bright lamellas (for direct comparison see also Figure~\ref{fig:EDS_10kK}). The bright area, reaching approximately into the center of the figure, is a roll-up of the sample. The minimum lamellar spacing is $\lambda \approx 32$ nm in the pristine sample. It should be noted that this value is the peak to peak value, analogous to a wavelength.

\subsection{Spheroidization and coarsening of lamellar structures\label{sec:res_spheroidization}}

{The spheroidization behavior for annealing at 300 $^\text{o}$C for 3 min of the pristine material can be observed in Figure~\ref{fig:exp_1_intro} (see also~[dataset]\cite{dumitraschkewitz.data.2020} video 1). In the pristine material, nucleation of polyhedra grains at the front of the colonies of the $\theta$ lamellas is observed, which grow at the cost of dissolving lamellas and form a mostly inter-connected $\theta$-grain network.}

\begin{figure}
    \centering
    \includegraphics[width=8cm]{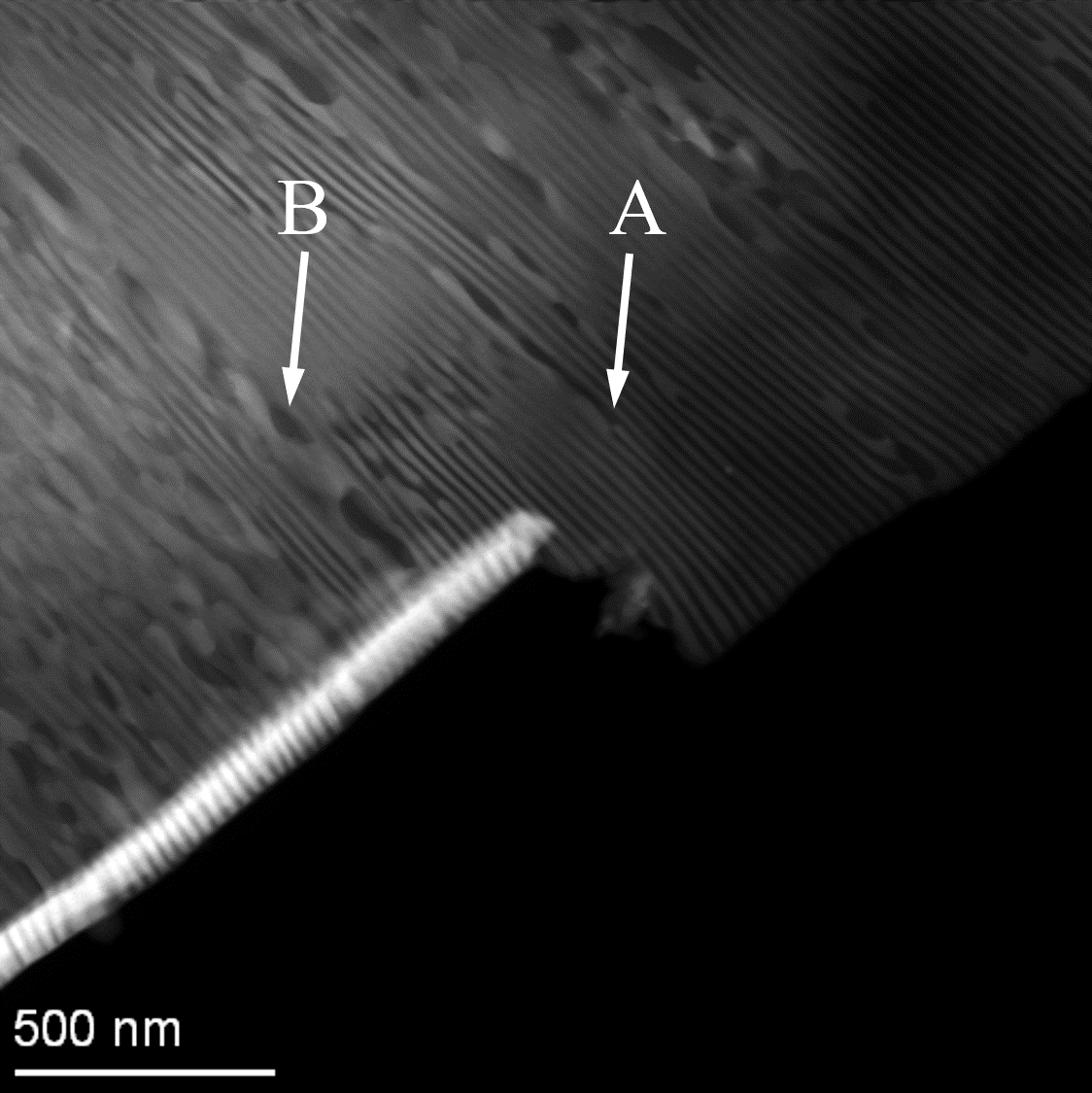}
    \caption{HAADF image of coarsening at 300 $^o$C (Video 2). Feature A shows thickening of a lamella. Feature B shows beginning spheroidization of lamellas.}
    \label{fig:lamella_movement}
\end{figure}

For the unidirectionally oriented lamella morphology (Figure~\ref{fig:exp_12_intro}), growth in width of $\theta$ lamellas (further refered to as ''thickening``, feature A in Figure~\ref{fig:lamella_movement}) and spheroidization (feature B in Figure~\ref{fig:lamella_movement}) is observed while annealing at 300 $^\text{o}$C.

\begin{figure}
    \centering
        \begin{subfigure}[t]{8.85cm}
            \centering
            \includegraphics[width=8cm]{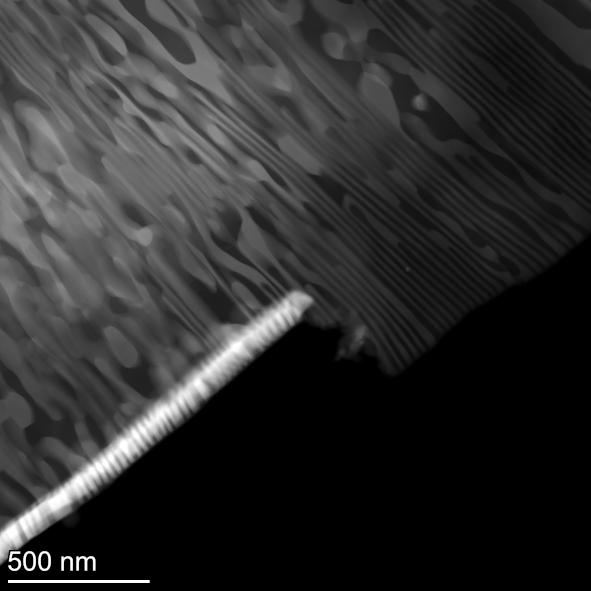}
            \caption{Coarsened at 300 $^o$C for 3 min}
            \label{fig:exp_14_main}
        \end{subfigure}
        \begin{subfigure}[t]{8.85cm}
            \centering
            \includegraphics[width=8cm]{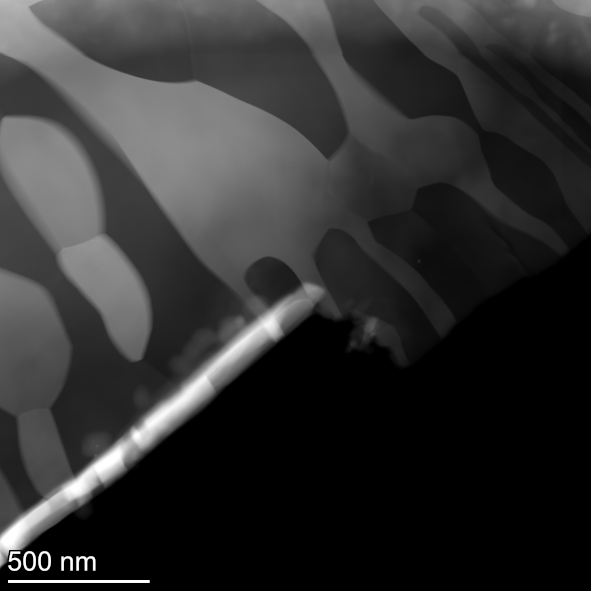}
            \caption{Coarsened at 500 $^o$C for 3 min}
            \label{fig:exp_17_main}
        \end{subfigure}
    \caption{HAADF images of coarsened material after pulse heating and free cooling. a) lamellar spheroidization and lamellar thickening. b) Additional coarsening heat treatment lead to coalescence and growth of grains.}
    \label{fig:exp_14_17}
\end{figure}

Several previously parallel lamellas are seen to form interconnected, significantly elongated round grains after 3 min at 300 $^\text{o}$C (Figure~\ref{fig:exp_14_main}).

Lamella faults are often found to be nucleation points for thickening and spheroidization. For thickening of lamellas, the increase in width follows the main direction of the lamella at the cost of neighboring lamellas (see~[dataset]\cite{dumitraschkewitz.data.2020}, video 2 and Figure~\ref{fig:lamella_movement}). Annealing at 500 $^\text{o}$C for 3 min leads to coarser elongated spheroidized grains (Figure~\ref{fig:exp_17_main}).

\subsubsection{Recrystallization kinetics of lamellar colonies}

The thickening of a lamellar at the cost of a neighboring lamella, feature A in Figure~\ref{fig:lamella_movement}, is measured to be $\approx$ 41 nm/s on average for an isothermal holding temperature of 300 $^o$C (see also~[dataset]\cite{dumitraschkewitz.data.2020}, video 2).

For several frames, the length of the initial (dissolving) lamella is measured and tracked. The time $t$ is calculated with a frame-to-frame time $\Delta t$ of 706 ms, setting the time for the reference frame $n_\text{frame, initial}$ as origin to zero, and for further frames numbers $n_\text{frame}$ according to Equation~\ref{eq:time_setting}.

\begin{align}
    \label{eq:time_setting}
    t = \Delta t \left(n_\text{frame}-n_\text{frame, initial}\right)
\end{align}

The first frame shows a length for the inital lamella of 584.4 nm ($\Delta s_\text{max}$) and the $\lambda$ value is $\approx$ 22 nm. In the last used frame the initial lamella vanished ([dataset]\cite{dumitraschkewitz.data.2020}, video 2, frames 148-168). The average velocity $v_\text{avg}$ is calculated according to Equation~\ref{eq:avg_vel}, where $\Delta t_\text{max}=14.12$ s, yielding 41 nm/s.

\begin{align}
    \label{eq:avg_vel}
    v_\text{avg} =& \frac{\Delta s_\text{max}}{\Delta t_\text{max}}
\end{align}

Interval velocities $v_\text{int}$ are computed according to Equation~\ref{eq:interval_vel}, where $\Delta s_\text{interval}$ and $\Delta t_\text{interval}$ are the respective length differences of the initial lamella and time differences between two frames.

\begin{equation}
    \label{eq:interval_vel}
    v_\text{int} = \frac{\Delta s_\text{interval}}{\Delta t_\text{interval}}
\end{equation}

The interval velocities vary from in the range of 23 - 71 nm/s.

\subsection{Melting and re-solidification\label{sec:res_melting}}

{Two re-solidification experiments are are presented in Figure~\ref{fig:resolidifications}. As the main parameter the cooling conditions are varied; pulse heating with free cooling and cooling with 100 K/s is utilized.}

During the progression in number of experiments, holes developed in the specimen at the partly covered membrane hole (Figure~\ref{fig:exp_25}, feature A). { It should be noted that the specimen is thicker at feature B). In general the sample area shrinks and the thickness can increase over the sequence of meltings.}

{Cooling with a rate of 100 K/s during solidification results in a coarse more 3-dimensional morphology (Figure~\ref{fig:exp_25} at feature B). Large thin dendritic crystals, either $\alpha$ or $\theta$, at the surface are covering lamella morphologies behind them. Not only a single structure through the whole thickness of the sample is apparent in the field-of-view.

Solidification after a pulse heating and free cooling resulted in unidirectionally oriented lamellas with a $\lambda_{\text{min}}$ of $\approx 22$ nm, see Figure~\ref{fig:first_remelt}.

\begin{figure*}
    \centering
    \begin{subfigure}{0.45\linewidth}
                \centering
                \includegraphics[height=8cm]{figures/HAADF_exp12_48kx}
                \caption{ pulse heating and free cooling}
                \label{fig:first_remelt}
    \end{subfigure}
    \begin{subfigure}{0.45\linewidth}
    \centering
            \includegraphics[height=8cm]{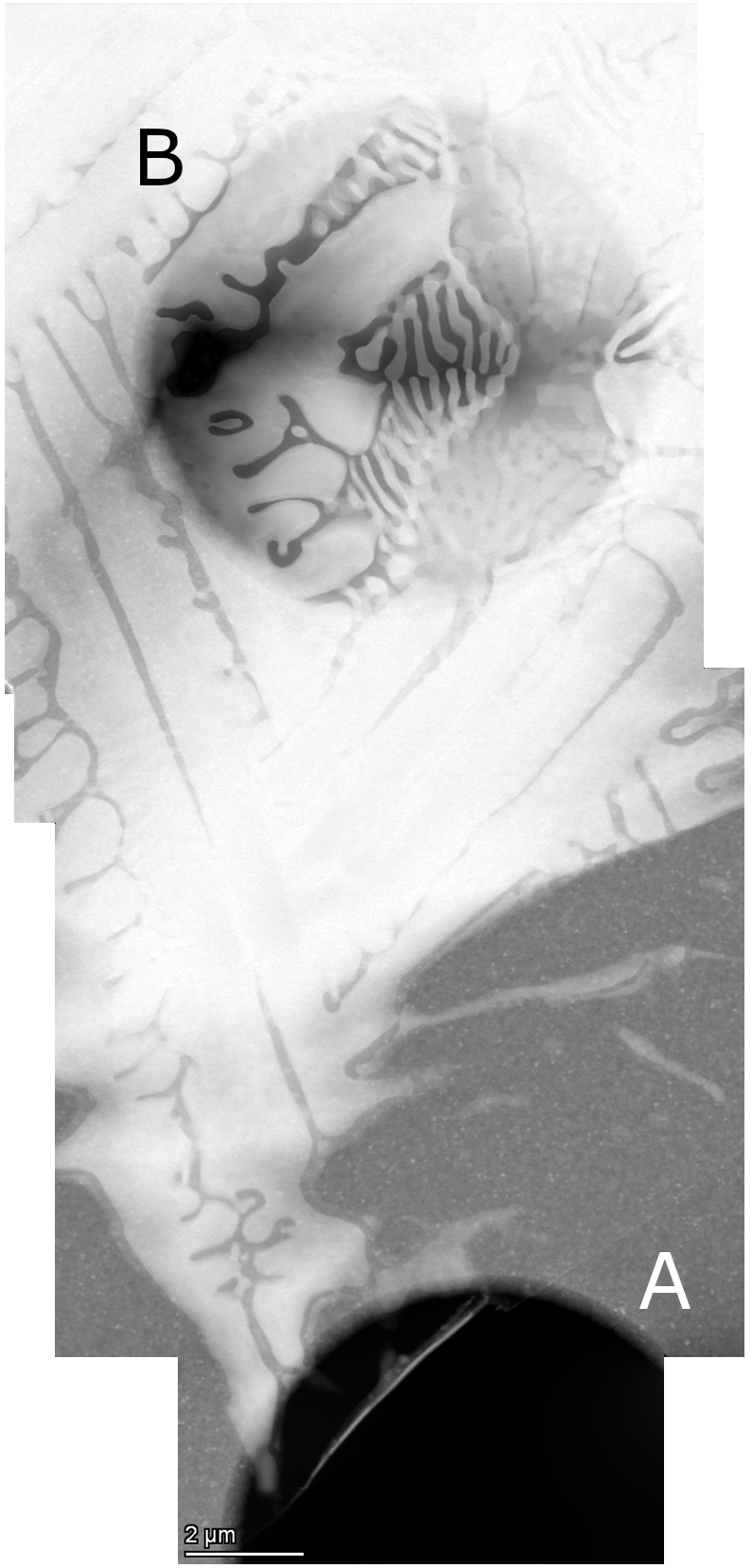}
            \caption{100 K/s}
           \label{fig:exp_25}
    \end{subfigure}
    \caption{HAADF images for re-solidification experiments with varied cooling conditions; after a pulse heating and free cooling (a) and cooling with 100 K/s (b). Image in a) was taken at the partly covered membrane hole; feature A in b). Following microstructures are obtained:  a) strong unidirectional lamellar, b) coarse lamellar and surface grains. Note that b) is a stitch out of 4 images from a video sequence.}
    \label{fig:resolidifications}
\end{figure*}

The overall composition of the pristine sample is measured to be $\approx 14.6$ at.\% Cu and $85.4$ at.\% Al in balance. Further EDS measurements are reported in Table~\ref{tab:chemical_analysis} and discussed in section~\ref{sec:comp_discussion}.

Notably, it should be mentioned that the roll-up feature of the image is retained, even after melting and re-solidifcation.}

\section{Discussion}
In the following, spheroidization is discussed and a term for the thermodynamical driving force for a special case is developed (Section~\ref{sec:disc_sphere}). Please note that additional details on the derivation of the model are given in the appendix and only the essential parts are reproduced in the main text for better readability and to focus on the main result. { Moreover, the re-solidified morphologies are discussed in Section~\ref{sec:disc_solidification}.}

\subsection{Lamella spheroidization and coarsening\label{sec:disc_sphere}}
A rough estimation of the coarsening rate of eutectic lamellas has already been conducted by Lemaignan~\cite{Lemaignan.1981}, but also stated that the geometrical situation is different for lamellar eutectic systems than in the applied model. The equation initially used for the estimation was built for smaller solid particles/dispersoids dissolving to benefit the growth of larger particles in a liquid~\cite{G.W.Greenwood.1956}.

In Reference~\cite{Gottstein.2007}, the driving force of recrystallization $p$ is given by the free enthalpy reduction $-dG$ gained by passing the grain boundary over a volume $dV$, see Equation~\ref{eq:driving_Gibbs}.

\begin{equation}
    \label{eq:driving_Gibbs}
    p = -\frac{dG}{dV}
\end{equation}

Several expressions for $p$ can be found assuming different driving forces for recrystallization, e.g. for continuous recrystallization  Equation~\ref{eq:recrystallization} is stated~\cite{Gottstein.2007}, where $\gamma$ denotes the grain boundary energy and $R$ the radius of curvature.

\begin{align}
\label{eq:recrystallization}
    p &= \frac{2\gamma}{R}
\end{align}

For the special case of thickening a $\theta$-lamella at the cost of a neighbor (Figure~\ref{fig:lamella_movement}, feature A), as observed in video 2 ([dataset]\cite{dumitraschkewitz.data.2020}) during annealing at 300 $^o$C, an expression for $p$ is developed (see~\ref{sec:thickening_model}) and reported in Equation~\ref{eq:drive_lamella}. $\gamma_{\alpha\theta}$ denotes the columnar interface energy of the $\alpha$/$\theta$ lamellas and $\lambda_\text{init}$ the initial lamella distance.

\begin{align}
\label{eq:drive_lamella}
    p &= \frac{2\gamma_{\alpha\theta}}{\lambda_\text{init}}
\end{align}

For a hemispherical grain ending in a differently oriented grain, Equation~\ref{eq:hemisphere} is given~\cite{GottsteinG.MolodovD.A.ShvindlermanL.S..1998}.
For a similar case, a triple junction, where a grain ends between two differently oriented grains, a driving force as in Equation~\ref{eq:equil_angle} is reported~\cite{M.UpmanyuD.J.SrolovitzL.S.ShvindlermanG.Gottstein.2002,GottsteinGSursaevaVShvindlermanLS..1999}, where $w$ represents the grain width. $\beta$ is the grain boundary angle with possible values ranging from 0 to $\pi/3$ in the model, if the shrinking grain is not dragged by the triple junction itself. 

Using $\lambda$ for $w$ shows that Equation~\ref{eq:drive_lamella} matches Equation~\ref{eq:hemisphere}, or respectively lies in the range of Equation~\ref{eq:equil_angle}, below the limiting case. If no drag of the triple junction slows the boundary movements, the angle $\beta = \pi/3$ as limiting case can be used~\cite{GottsteinG.MolodovD.A.ShvindlermanL.S..1998}; comparing Equation~\ref{eq:drive_lamella} or~\ref{eq:hemisphere} to~\ref{eq:equil_angle} leads a ratio of 2 to $2\pi/3$. It should be noted that Equation~\ref{eq:equil_angle} was derived by a formalism surface tension acting on boundary~\cite{GottsteinGSursaevaVShvindlermanLS..1999}. 

\begin{align}
     \label{eq:hemisphere}
     p &= \frac{2\gamma}{w}\\
    \label{eq:equil_angle}
     p &= \frac{2\beta\gamma}{w}
\end{align}

\begin{figure}
    \centering
    \includegraphics[width=5.5cm]{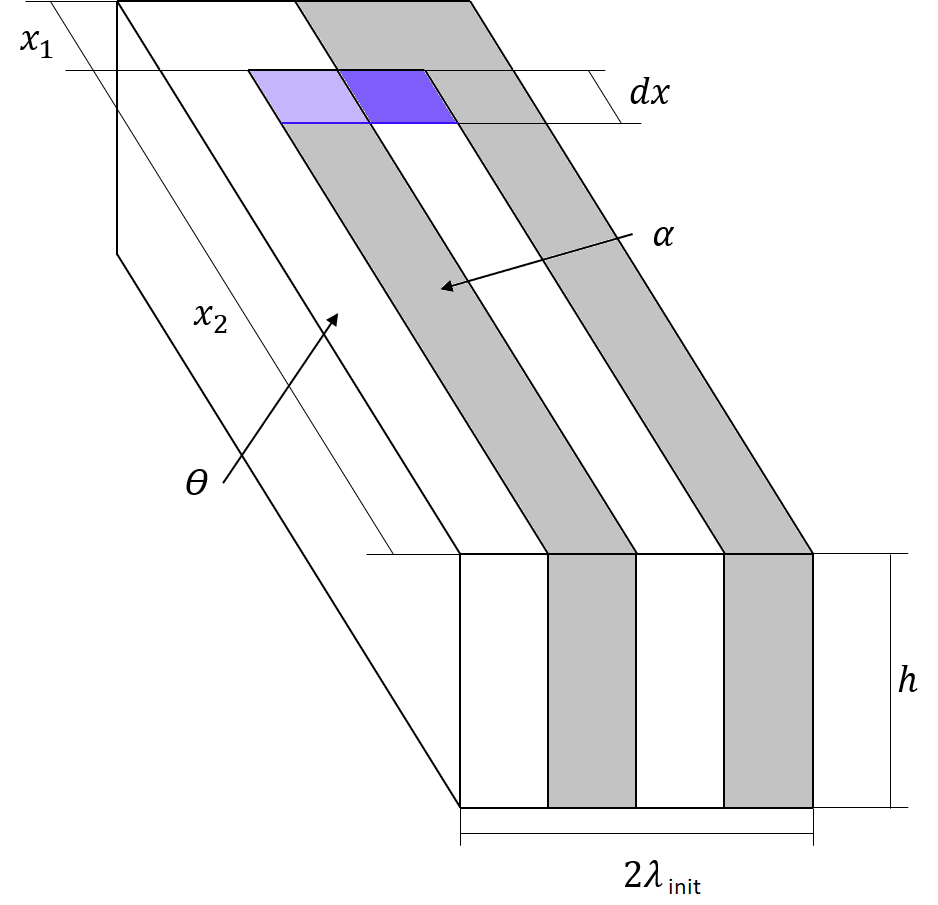}
    \caption{Simplified geometry for lamella thickening. $h$ is the thickness of the sample, $dx$ is the movement of the interface with a time interval and $\lambda_\text{init}$ is the initial lamella spacing. The bright blue area changes from $\alpha$ to $\theta$ phase, and the dark blue area from $\theta$ to $\alpha$, during the interface movement of $dx$.}
    \label{fig:simple_geometry}
\end{figure}

However, in the present case, four grains meet with two of them of different phase (Figure~\ref{fig:simple_geometry}). The different geometric situation will therefore contribute to differences in derived expressions.

The geometry of the connected grains cannot be clearly observed with the used magnification of Figure~\ref{fig:lamella_movement}. In the derivation of Equation~\ref{eq:drive_lamella}, the influence of bowed interface lines is neglected. Flat lines would actually create locking of the boundary lines~\cite{GOTTSTEIN.2005}, but, the free ending corners in Figure~\ref{fig:simple_geometry} introduce points of instability. Here it should be emphasized that the lamella defects are seen to be nucleation points for thickening of lamellas and also elicit spheroidization. Further, for the energetic derivation formalism only the self-similarity of the front interface before and after the movement of $dx$ is needed. 

Another general simplification made for Equation~\ref{eq:drive_lamella}, as in the other two dimensional cases, is the assumption of a constant height of the sample, when actually the sample is likely wedge-shaped and gets thinner towards the sample edge. This could explain the faster movement~\cite{bauer1990mechanisms,G.GottsteinandL.S.Shvindlerman.1992} of feature A in Figure~\ref{fig:lamella_movement}, in comparison to a slow moving, similar feature in the opposite direction. A deleterious effect of grain boundary grooving is not expected for Al due to an existing oxide layer~\cite{G.GottsteinandL.S.Shvindlerman.1992,Zhang.2011}. Faster recrystallization velocity could be caused by smaller sample thickness.

An estimation of the driving force acting in Figure~\ref{fig:lamella_movement} by Equation~\ref{eq:drive_lamella}, with $\gamma_{\alpha\theta}\approx253$ mJ/m$^2$~\cite{Kokotin.2014} and $\lambda_\text{init}\approx22$ nm, gives a value of $\approx$ 20 MPa, reaching driving forces in the order of primary recrystallization of heavily cold worked metals~\cite{Gottstein.2007}. 

To physically assess the kinetics of lamellar thickening, we first consider the thickening rate without taking into account the exact mechanism. With the driving force $p$ and the mobility $m$ of the phase boundaries, the velocity $\Tilde{v}$ according to Gottstein~\cite{Gottstein.2007} results in

\begin{align}
    \label{eq:mobility}
    \Tilde{v} &= mp,
\end{align}
whereby the mobility is specified in relation to the migration-determining diffusion coefficient $D_m$, the jumping distance $b$, the Boltzmann constant $k$ and the absolute temperature $T$ with

\begin{align}
    \label{eq:nernst_einstein_gottstein}
    m &= \frac{b^2D_m}{kT}.
\end{align}

With $a=0.405$ nm as the fcc lattice constant, $b$ is estimated by $b = \sqrt[3]{a^3/4}$~\cite{T.E.VOLINandR.W.BALLUFE.1968}. The diffusion coefficient can therefore be expressed by Equation~\ref{eq:diffusion_result}.

\begin{align}
\label{eq:diffusion_result}
    D_m = kT \frac{\lambda_\text{init}}{2\gamma_{\alpha\theta} b^2}\Tilde{v}
\end{align}

Inserting the isothermal temperature of 300 \textsuperscript{o}C (573 K), with the values for $\lambda_\text{init}$ and $\gamma_{\alpha\theta}$ given earlier in the text, and varying $\Tilde{v}$ from 23 to 71 nm/s results in values for $D_m = 1.22 - 3.75 \times 10^{-12}$ cm$^2$/s. Literature data for self-diffusion of Al are at comparable values of $D_\text{Al} = 5.32 \times 10^{-13}$ cm$^2$/s measured by void annihilation~\cite{T.E.VOLINandR.W.BALLUFE.1968} and $D_\text{Al} = 1.85 \times 10^{-13}$ cm$^2$/s determined by tracer experiments~\cite{lundy1962diffusion}. For Cu volume diffusion in Al~\cite{anand1965diffusion}, a similar value of $D_\text{Cu} = 4.65 \times 10^{-13}$ cm$^2$/s is found, see also Table~\ref{tab:diffusion_constants}. Considering that diffusion is expected to be faster along interfaces, values are in reasonable agreement, but should not be over-interpreted. In this context, it is important to note that the thickening of a $\theta$-lamella takes place at the expense of neighbor lamella, as observed in video 2, and is achieved by a movement of the phase boundary perpendicular to the thickening direction. Energy criteria, as in Equation~\ref{eq:recrystallization}, have been developed for single phase materials, but for multiphase materials more effects would need consideration, e.g. the combined diffusion of Al and Cu, solute drag of additional elements~\cite{VERHASSELT1999887} and the orientation dependence of the $\gamma_{\alpha\theta}$ interface energy~\cite{Kokotin.2014}. A sound description is anything but trivial and is out of the scope of this paper.

Modeling methods as phase-field~\cite{Pinomaa.2020} or molecular dynamics~\cite{M.UpmanyuD.J.SrolovitzL.S.ShvindlermanG.Gottstein.2002} simulations could deepen the understanding of the observed phenomenom.

An important general observation which should be pointed out is that in comparison to the pristine material (Figure~\ref{fig:pristine_intro}), the strong uni-directional morphology (Figure~\ref{fig:exp_12_intro}) tends to have a higher resistance to recrystallization, compare Figure~\ref{fig:exp_14_main} to Figure~\ref{fig:exp_1_intro} which both experienced annealing for 3 min at 300 $^o$C. This results from fewer lamella faults, which act as nucleation points for recrystallization, in the strong uni-directional morphology.

\subsection{Re-solidified morphologies\label{sec:disc_solidification}}
{
A small lamella spacing, according to eutectic scaling laws (Equation~\ref{eq:tiller_2}) has been expected for rapid cooling conditions.

Rapid surface re-solidification experiments showed that lamellar spacing can only reach a minimum of $\approx$ 17 nm and increases again with further increasing solidification velocity~\cite{M.ZimmermannM.CarrardW.Kurz.1989}. When a critical solidification speed is reached a cellular and dendritic microstructure is expected for eutectic composition~\cite{S.C.GILLW.KURZ.1993}. Further a} phase replacement for the {$\theta$ phase} has been reported for solidification velocities where the regular eutectic morphology breaks down, and a re-increase in $\lambda$ spacing is observed~\cite{S.C.GILLW.KURZ.1993,M.ZimmermannM.CarrardW.Kurz.1989}. {For even higher solidification velocities so-called banded regions~\cite{S.C.GILLW.KURZ.1993, Bathula.2020} which appear at very high solidification velocities and are the result of an oscillatory solidification.} Here no $\alpha/\theta$ phase constituents are present any more, but a partitionless (up to a resolution of about 3 nm) solidification of $\alpha$ alternating to $\theta^\prime$ phase, which has a kinetic advantage for nucleation due to coherent interfaces to $\alpha$.~\cite{Bathula.2020, Zweiacker.2018}

{However, in general not only the solidification velocity determines the morphology, but also the temperature gradient and the melt composition during the course of solidification.~\cite{Kurz.1979,Kurz.2019}

Nucleation~\cite{Simon.2017, Yang.2021} will influence the amount of undercooling and together with the external heat exchange the recalescence~\cite{C.G.LeviR.Mehrabian.1982} behavior.

The conduction of heat to the unheated parts of the chip is the dominant effect for cooling the membrane. The heat exchange from the sample to the chip will not be ideal heat conduction, but better described by a thermal contact conductance~\cite{Minakov.2020}, i.e. a heat exchange coefficient between the chip and the solid or liquid specimen. Average cooling rates for the modeling of splat-cooling are discussed in Ref.~\cite{ROBERTC.RUHL.1967} for both mentioned heat exchange cases and varied thickness.

Comparing to Ref.~\cite{McKeown.2014}, this work's fastest applied cooling (via pulse heating and free cooling) is estimated to be slower. Due to lack of high sampling frequency, the actual temperature of the membrane was not followed with high enough temporal resolution for the pulse heating with free cooling experiment. The chip membrane part which is heated via Joule heating~\cite{Allard.2009} here is $\approx 200\times 200$ $\mu$m$^2$ (see Figure~\ref{fig:sketch_membrane}) comparing to the meltpoolsize of $\approx27\times35$ $\mu$m$^2$ in Ref.~\cite{McKeown.2014}. Still, rapid coooling conditions can be realized, for the predecessor chips membrane a maximum cooling rate in the order of $10^6$ K/s is reported in Ref.~\cite{Allard.2009}.

The area of the sample is seen to decrease with increasing time in the melt/subsequent number of melting and re-solidification cycles, compare feature A in Fig.~\ref{fig:first_remelt} to Fig.~\ref{fig:exp_25}. With decreasing area the thickness of the specimen increases, see also section~\ref{sec:sample_shape}.}

Direct comparison to the microstructure selection map from e.g. Ref.~\cite{S.C.GILLW.KURZ.1993} is difficult; the solidification structures are produced upon an imposed cooling rate at the {chip}, but the selection map requires a solidification {velocity, which is indirectly deduced from the laser movement velocity~\cite{M.ZimmermannM.CarrardW.Kurz.1989}.

For the pulse heating experiment with free cooling (Fig.~\ref{fig:first_remelt}) a lamellar morphology with a spacing of $\approx 22$ nm is observed. We could identify the Al$_2$Cu lamella for a pulse heating experiment as $\theta$-Al$_2$Cu (see Supplementary Material Fig.1). Surface grains in a dendrite morphology and a partial lamellar coarse structure are observed with a cooling rate of 100 K/s. The change of morphology from one to another is attributed to the imposed cooling rate at the chip and change in thickness of the specimen.}

The exact linkage between the quantities {as temperature gradient, cooling rate and solidification velocity} depends on the solution of the \textit{Stefan} problem~\cite{Glicksman.2011, H.S.CarslawandJ.C.Jaeger.1959} {(moving boundary condition). The temperature distribution could be simulated e.g. via the finite difference method~\cite{J.Crank.1975, ROBERTC.RUHL.1967}, finite element method (e.g. as in Ref.~\cite{Keller.2017}) or other numerical methods.}

It should be noted that in general the occurrence of additional elements~\cite{Hecht.2004, Chanda.2020} will strongly affect the solidification behavior. Therefore ternary/off-eutectic compositions could limit the applicability of the calculated~\cite{S.C.GILLW.KURZ.1995} microstructure selection map of Reference~\cite{S.C.GILLW.KURZ.1993}.

\subsubsection{Composition analysis\label{sec:comp_discussion}}

{Looking at Al and Cu contents only (Table~\ref{tab:chemical_analysis}), one observes an off-stoichiometry for the pristine state of the $\theta$-phase (Al$_2$Cu), comparing an average Cu content of 26.3 to 33.3 at.\% Cu. Calculating the ratio of the overall sample composition (14.6 at.\% Cu) to the expected (OES measured) value of 17.5 gives a value of 0.83, proceeding the same way for the Cu content of the Al$_2$Cu phase gives a ratio of 0.79.} The measured deficiency is therefore likely caused by an underestimation of the Cu content due to the used k-factor method. Besides the Al and Cu signal expected from the specimen, further artefact element signals are detected as discussed in~\ref{sec:EDS_results}, which could additionally contribute to uncertainty in EDS composition. Furthermore, for very small sized features, EDS measured values (e.g. fine lamellas) could be influenced by limited resolution of the sampling area. No significant correlation between different morphologies and Cu content are deduced from EDS.

\section{Summary and conclusions\label{sec:summary}}
With this set of experiments it is demonstrated that in-situ STEM solidification of an nanoscaled eutectic alloy is possible, using a recently developed sample preparation method~\cite{Tunes.2020} and a MEMS based heating/cooling holder. { Even subsequent experiments with the same specimen are conducted.

Two examples for potential investigations with the newly developed methodology are reported in detail, influence of the cooling conditions on the rapid solidification morphology and recrystallization heat treatments.

With application of pulse heating and free cooling from the melt, a strong uni-directional, nanostructured morphology could be observed, reaching a lamella spacing of 22 nm. Using a cooling rate of 100 K/s, the re-solidified morphology is coarser by at least an order of magnitude.}

Analysis of in-situ recrystallization experiments shows an average interface velocity of 41 nm/s at 300 $^o$C for lamella thickening. For this special recrystallization case a term for the thermodynamical driving force is developed (Equation~\ref{eq:drive_lamella}). Besides this findings a general higher resistance of strongly oriented lamellas against recrystallization is observed.

{Applying the demonstrated experimental setup opens up a new way for in-situ solidification studies of Al based alloys and likely other metallic materials.} Post-solidification heat treatments, like additional heatspikes observed after solidification, as for e.g. mimic additive manufacturing, are able to be investigated. Although some limitations due to free surface of the TEM specimen should be considered~\cite{Saidi.2021, Dong.2019, Dumitraschkewitz.2019}.

\section{Declaration of competing interest}
The authors declare no competing interests.
\section{Acknowledgements}
This research was supported by funding from the European Research Council (ERC) under the European Union’s Horizon 2020 research and innovation program (grant No. 757961). {The transmission electron-microscopy facility used in this work received funding from the Austrian Research Promotion Agency (FFG) project known as ``3DnanoAnalytics'' under contract number FFG-No. 858040.}

The authors thank Prof. J\"urgen Eckert for the possibility of using the meltspinning facility at the Erich Schmid Institute. Mr. Stemper is kindly thanked for the help and introduction to the induction furnace. Mr. Cattini is kindly thanked for SEM investigations. Ms. Tatzreiter's help with metallography is very much acknowledged.

\bibliography{AlCu_upper_case}

\clearpage
\appendix
\section{Model for lamellar thickening\label{sec:thickening_model}}
A simplified geometry as seen in Figure~\ref{fig:simple_geometry} is used. Due to simplification, the three-dimensional problem is reduced to a two-dimensional problem. Instead of interface areas only the interface 'lines' need to be counted.

Before the movement of the boundary the interface lines have the length $s_0$ seen in Equation~\ref{eq:mov_before} and after the movement of $dx$ the length $s_1$ of Equation~\ref{eq:move_after}. $d_\alpha$ and $d_\theta$ are the respective width of the $\alpha$ and $\theta$ lamella and $x_1$, $x_2$ the length of the thickened, respectively the initial lamella. This interface length change of Equation~\ref{eq:delta_s} over the volume $\lambda h dx$, inserted into Equation~\ref{eq:driving_Gibbs}, leads to Equation~\ref{eq:driving_force}.

\begin{align}
    \label{eq:mov_before}
    s_0&=x_1+d_\alpha+x_2 + 2x_2+d_\theta\\
    \label{eq:move_after}
    s_1&=x_1+dx+d_\alpha+(x_2-dx) + 2(x_2-dx)+d_\theta\\
    \label{eq:delta_s}
    s_1-s_0 &= -2dx\\
    \label{eq:driving_force}
    p &= -\frac{-2  h \gamma_{\alpha\theta} dx}{\lambda_\text{init} h dx}
\end{align}
Therefore the energy consideration leads to an expression as in Equation~\ref{eq:drive_lamella} during interface/grain boundary movement.

In Table~\ref{tab:diffusion_constants} information is given, which is used for calculation of the literature diffusion constants in section~\ref{sec:disc_sphere}. {The interested reader is referred to Reference~\cite{Dahlborg.2007} for diffusion constants in the liquid state.}

\begin{table}
\centering
\caption{\label{tab:diffusion_constants} Numerical values for diffusion constant calculations according to $D=D_0\exp\left(-\frac{Q}{RT}\right)$.}
\begin{threeparttable}
\begin{tabularx}{8.85cm}{>{\raggedleft\arraybackslash}X>{\raggedleft\arraybackslash}X>{\centering\arraybackslash}X>{\raggedleft\arraybackslash}X}
\hline\hline
Ref.& $D_0$ [cm$^2$/s] & Q [kJ/mol]\\
\hline
\cite{T.E.VOLINandR.W.BALLUFE.1968}& 0.176& 126.39\\
\cite{lundy1962diffusion}& 1.710& 142.29\\
\cite{anand1965diffusion}& 0.150& 126.40\\
\hline\hline
\end{tabularx}
\end{threeparttable}
\end{table}

\section{EDS analysis, Contamination, oxide layer and the occurence of Si\label{sec:EDS_results}}

\begin{figure}
    \centering
    \begin{subfigure}{8.85cm}
        \centering
        \begin{subfigure}{3.98cm}
            \includegraphics[width=\linewidth]{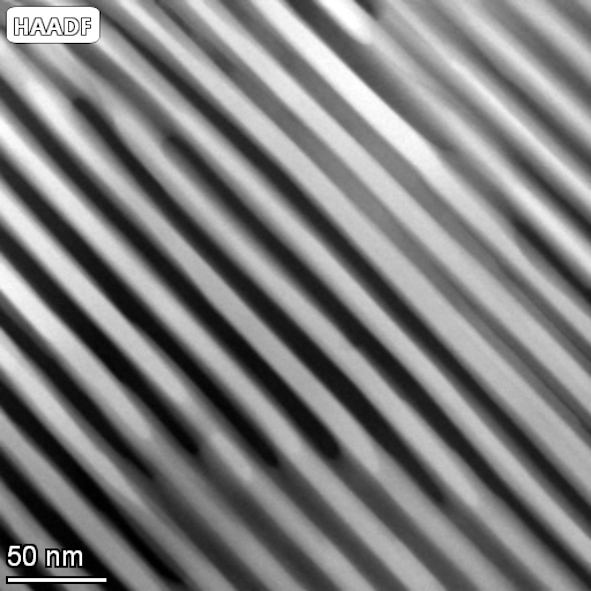}
            \caption{HAADF}
        \end{subfigure}
        \begin{subfigure}{3.98cm}
            \includegraphics[width=\linewidth]{figures/EDS_10kK_190kx}
            \caption{Al (gren), Cu (red)}
        \end{subfigure}
    \end{subfigure}
    \begin{subfigure}{8.85cm}
        \centering
        \begin{subfigure}{3.98cm}
            \includegraphics[width=\linewidth]{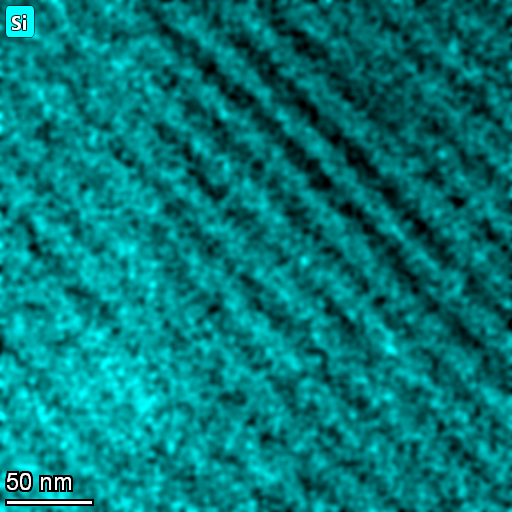}
            \caption{Si}
        \end{subfigure}
        \begin{subfigure}{3.98cm}
            \includegraphics[width=\linewidth]{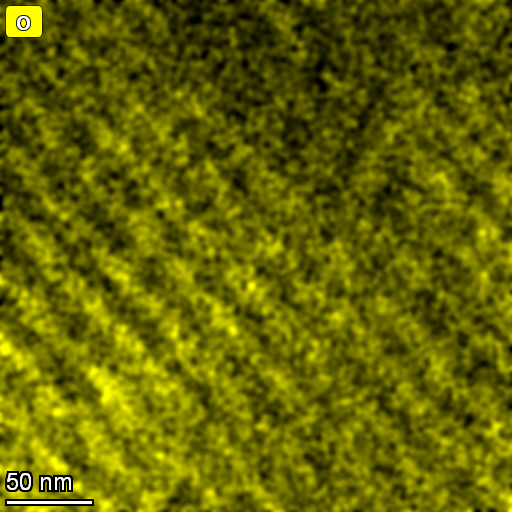}
            \caption{O}
            \label{fig:oxygen}
        \end{subfigure}
    \end{subfigure}
    \caption{HAADF and EDS images after of uni-directional lamellas (Figure~\ref{fig:first_remelt}). Bright areas are Cu rich ($\theta$-Al$_2$Cu) and dark areas are rich in Al ($\alpha$-Al). It can be seen that some lamellas are likely interconnected over a small surface layer of the respective phase. Si seems to be partitioned into $\theta$-Al$_2$Cu, while O is more prominent in $\alpha$-Al lamellas.}
    \label{fig:EDS_10kK}
\end{figure}

{Compositional analysis is performed via EDS as reported in Table~\ref{tab:chemical_analysis}.} In Figure~\ref{fig:EDS_10kK} the uni-directional lamellar structure (Figure~\ref{fig:first_remelt}) is shown with the respective EDS mappings for Al, Cu, Si and O. The Cu rich lamellas show approximately the composition of $\theta$-Al$_2$Cu, and Al rich show the composition expected for $\alpha$-Al with solute Cu. Despite clearly discernible lamellas for Al and Cu, there seem to be inter-connected lamellas, possibly over layers of surface crystals of the respective phase. The resulting Cu and Al contents of the by HAADF contrast discernible phases are reported in Table~\ref{tab:chemical_analysis}.

As seen in Figure~\ref{fig:EDS_10kK}, Si seems to be partitioned into the $\theta$-Al$_2$Cu lamellas, while O is prominent in $\alpha$-Al lamellas. The same partitioning behavior is also observed for recrystallized structures. No partitioning could be observed for C. 

\begin{table}

\centering
\caption{\label{tab:chemical_analysis} EDS chemical analysis. For evaluation Al, Cu as possible elements were chosen. C and O as possible contamination and further Si signal (originating from the Si$_3$N$_4$ membrane) is neglected in the evaluation.}
\begin{threeparttable}
\begin{tabularx}{8.85cm}{>{\raggedleft\arraybackslash}p{0.7cm}>{\centering\arraybackslash}X>{\centering\arraybackslash}X>{\raggedleft\arraybackslash}p{0.7cm}>{\raggedleft\arraybackslash}p{0.7cm}}
\hline\hline
sample state& morphology& HAADF phase& Al [at.\%]&Cu [at.\%]\\
\hline
pristine & lamellar& bright\footnotemark[1]& 73.7& 26.3\\
 & & dark\footnotemark[2]& 96.6& 3.4\\
 & & overall\footnotemark[3]& 85.4& 14.6\\
\hline
pulse heated & lamellar& bright& 73.9& 26.1\\
 & & dark& 98.7& 1.3\\
 & & overall& 86.6& 13.4\\
\hline
cooled with 100 K/s\footnotemark[4] & dendritic surface grains, coarse lamellas& & & \\
\hline\hline
\end{tabularx}
\begin{tablenotes}
\item [1]{Sampling area of $\theta$-Al$_\text{2}$Cu.}
\item [2]{Sampling area of $\alpha$-Al.}
\item [3]{Sampling area of overall morphology.}
\item [4]{No EDS measured.}
\end{tablenotes}
\end{threeparttable}
\end{table}

C is a typical surface contaminant and likely emerges from cleaning residues. No defined aggregation behavior in the sample has been observed.

O is expected from surface oxidation, but also partitioning seems to be apparent into $\alpha$-Al. While higher surface oxidation of $\alpha$-Al phase in the pristine sample could be possible, for the re-solidified sample (as in Figure~\ref{fig:EDS_10kK}), it is not expected for newly formed lamellas, due to operation under {high vacuum (HV)} conditions in the STEM. The amount measured (order of percentage) is far beyond solubility for an interstitial element, especially for Al with almost non-existent solubility of O. In fact, for the shown state in Figure~\ref{fig:oxygen}, only little partitioning, approx. 0.5 \% at. absolute excess, is observed. The found intensity is likely overlaid with an existing signal from an oxide layer, and possibly an artefact of stray radiation.

Si is observed to be partitioned into $\theta$-Al$_2$Cu. A source of Si signal can be stray signal from the Si$_3$N$_4$ holder. Direct signal from the membrane can be excluded due to observation at a membrane hole. The observed supposedly partitioning of Si into the $\theta$-Al$_2$Cu phase can be explained by the x-ray fluorescence of excited Cu atoms. Si in the material is further ruled out by an additional SEM EDS of a ribbon from the same batch, showing only Al and Cu, but no Si signals.

\subsection{Sample shape\label{sec:sample_shape}}

After the sample preparation, the sample is thought to be in a wedge shape, a form with varying thickness over sample in plane dimensions, but with a low thickness to width ratio ($\approx 1/500$) in general. An oxide layer is expected at the surface due to preparation at atmosphere and rapid (passivating) surface oxidation of Al alloys.

If the initial contact-angle $\psi$ between liquid (sample) and solid (membrane) is lower than the equilibrium value determined by Youngs' equation (Equation~\ref{eq:young}), then during the liquid state an increase in thickness by spheroidization{/balling} is expected until the equilibrium angle and shape is reached. $\sigma_{sg}$, $\sigma_{sl}$ and $\sigma_{lg}$ here denote respectively the solid-gas, solid-liquid and the liquid-gas surface tensions. Due to the low aspect ratio, the assumption of a too low initial contact angle seems plausible. The spheroidization to reach equilibrium shape is in general a time and temperature dependent process, due to needed directional movement, or flow, of atoms by diffusion in the melt.

\begin{equation}
\label{eq:young}
\cos(\psi) = \frac{\sigma_{sg}-\sigma_{sl}}{\sigma_{lg}}
\end{equation}

{It should be noted that other metal systems than Al based systems might exhibit different spheroidization/balling characteristics.}

\subsection{Oxide layer and pile-up}

The role of a surface oxide layer at the specimen between the sample and holder is difficult to judge. The oxide layer could contribute to a decreased thermal contact conductance coefficient. An oxide layer formed at atmosphere is usually only some few nanometers in thickness and of amorphous nature. Cracking surface oxide layers upon heating could be expected, likely due to different heat expansion coefficients of oxide and metal. However, the remaining roll-up artefact after several melting and re-solidification events of the sample hints to some form stability of at least some parts of the oxide layer. 

\end{document}